\documentclass[12pt,preprint]{aastex}

\newcommand{\ms}{m\,s$^{-1}$}
\newcommand{\kms}{km\,s$^{-1}$}
\newcommand{\msun}{${\rm M}_{\sun}$}
\newcommand{\mathmsun}{{\rm M}_{\sun}}

\newcommand{\ELODIE}{{\footnotesize ELODIE}}
\newcommand{\HIRES}{{\footnotesize HIRES}}
\newcommand{\TODCOR}{{\footnotesize TODCOR}}
\newcommand{\OHP}{{\footnotesize OHP}}
\newcommand{\LTE}{{\footnotesize LTE}}

\newcommand{\hd}{{HD~178911}}
\newcommand{\hdb}{{HD~178911B}}
\newcommand{\hda}{{HD~178911A}}

\newcommand{\Teff}{${\rm T_{eff}}$}

\shorttitle{A Planet in the System HD~178911}
\shortauthors{Zucker et al.}

\begin{document}

\title{A Planet Candidate in the Stellar Triple System 
HD~178911\altaffilmark{1}\altaffiltext{1}{
Based on observations made at the Observatoire de 
Haute-Provence (French CNRS) and at the W.M.~Keck Observatory, which 
is operated as a scientific partnership among the Californian 
Institute of Technology, the University of California and the 
National Aeronautics and Space Administration. The Observatory was 
made possible by the generous financial support from the W.M.~Keck 
Foundation.}} 

\author{S. Zucker\altaffilmark{2},
D. Naef\altaffilmark{3},
D.W. Latham\altaffilmark{4},
M. Mayor\altaffilmark{3},
T. Mazeh\altaffilmark{2},
J.L. Beuzit\altaffilmark{5},
G. Drukier\altaffilmark{2,6},
C. Perrier-Bellet\altaffilmark{5},
D. Queloz\altaffilmark{3},
J.P. Sivan\altaffilmark{7},
G. Torres\altaffilmark{4},
and S. Udry\altaffilmark{3}}

\altaffiltext{2}{School of Physics and Astronomy, Raymond and Beverly
Sackler Faculty of Exact Sciences, Tel Aviv university, Tel Aviv 69978,
Israel}
\altaffiltext{3}{Observatoire de Gen\`{e}ve, 51 ch. des Maillettes,
CH-1290 Sauverny, Switzerland}
\altaffiltext{4}{Harvard-Smithsonian Center for Astrophysics, 
60 Garden Street, Cambridge, MA-02138, USA}
\altaffiltext{5}{Laboratoire d'Astrophysique, Observatoire de
Grenoble, Universit\'{e} J. Fourier, BP 53, F-38041 Grenoble, France}
\altaffiltext{6}{Present address: 
Department of Astronomy, Yale University, P.O.
Box 208101, New Haven, CT 06520-8101, USA}
\altaffiltext{7}{Observatoire de Haute-Provence, 
F-04870 St-Michel L'Observatoire, France}

\begin{abstract}

We report the detection of a low-mass companion orbiting the
solar-type star \hdb, the distant component of the stellar triple
system \hd. 
The variability of
\hdb\ was first detected using radial-velocity measurements obtained
with the \HIRES\ spectrograph mounted on the 10--m Keck\,1 telescope at
the W.M.~Keck Observatory (Hawaii, USA). We then started an intense
radial-velocity follow up of the star with the \ELODIE\ echelle
spectrograph mounted on the 1.93--m telescope at the Observatoire de
Haute-Provence (France) in order to derive its orbital solution. 
The detected planet candidate has an orbital period of 71.5 days and a
minimum mass of 6.3\,${\mathrm M_{\rm Jup}}$.  
We performed a spectral analysis of the star, which shows that the
lithium abundances in the system are similar to those in another known
planet-hosting wide binary stellar system, 16~Cyg. 
In both systems the lithium is undetected in the atmosphere of the visual
secondary harboring the planetary companion, but is easily detected in the
spectrum of the visual primary.
We discuss this
similarity and its ramifications.

\end{abstract}

\doublespace

\section{Introduction}

We report on radial-velocity measurements of \hdb, the fainter
component of the visual binary \hd\ (HR~7272, BD~+34\arcdeg\,3439, ADS
12101, HIP 94075/6), which reveal the presence of a 6.3\,Jovian-mass
companion (minimum mass). As the brighter star of the system --- \hda, is
actually a close binary (McAlister et al.\ 1987; Tokovinin et al.\
2000, hereafter T00), the discovery reported here is of a planet
candidate in a stellar triple system.

The variable velocity of \hdb\ was first noticed by the {\sl G--Dwarf
Planet Search} (Latham 2000). This was a reconnaissance monitoring of
nearly 1000 nearby G dwarfs with the \HIRES\ high-resolution
spectrograph (Vogt et al.\ 1994) mounted on the 10--m Keck~1 telescope
at the W.M.~Keck Observatory (Hawaii, USA) to identify extrasolar
planet candidates.  The star was then followed up by the {\sl \ELODIE\
Planet Search Survey} team (Mayor \& Queloz 1996, Udry, Mayor \&
Queloz 2001) using the \ELODIE\ fiber-fed echelle spectrograph
(Baranne et al.\ 1996) mounted on the Cassegrain focus of the 1.93--m
telescope at the Observatoire de Haute-Provence (CNRS, France).

The \ELODIE\ velocities were obtained by cross-correlating the observed
spectra with a numerical template.  The instrumental drifts were
monitored and corrected using the ``simultaneous Thorium-Argon technique''
with dual fibers (Baranne et al.\ 1996). The precision achieved with
this instrument for HD~178911B 
is of the order of 10\,m\,s$^{\rm -1}$. The \HIRES\
instrumental drifts were monitored using an Iodine gas
absorption cell (Marcy \& Butler 1992). The radial velocities were
derived from the spectra using the \TODCOR\ code (Zucker \& Mazeh 1994), a
two-dimensional correlation algorithm.

The first discovery resulting from the collaboration between our teams
was the determination of the orbital solution for 
HD~209458 (Mazeh et al.\ 2000) that led to the detection of the first
photometric (Charbonneau et al. 2000\footnote{The ingress of a
transit was also independently observed by Henry et al.\ 2000.}) and
spectroscopic (Queloz et al.\ 2000) transits of an extrasolar
planet. 
Recently we also detected the companion to HD~80606 and determined its
orbital solution. With e=0.93, it is the most eccentric planetary orbit
currently known (Naef et al.\ 2001).

Our observations of \hdb\ started in July 1998 with \HIRES. With a
velocity difference of 63\,\ms\ in about one month between the first two
measurements, it was tagged as a definite variable.  This was confirmed
by the next four observations in April and May 1999.  In July 1999, we
started an \ELODIE\ radial-velocity follow up of 6 non-active
slowly-rotating radial-velocity variable stars detected with \HIRES,
including \hdb. The combination of the two sets of radial velocities
enabled us to derive the orbital solution we
describe in Section \ref{RVs}. 

In Section \ref{stars} we consider the stellar characteristics of
the triple system. In particular we study the lithium abundances of the
system, specially because another wide binary known to harbor a planet
--- 16~Cyg --- exhibits a peculiar lithium abundance pattern. There the
lithium seems to be depleted in the fainter stellar companion hosting
the planet compared with the brighter stellar component. We find a
similar phenomenon in \hdb. Section \ref{lithium} discusses our
finding by reviewing the three models proposed to explain the lithium
abundance in 16~Cyg, and examines them in the context of \hdb.  Section
\ref{conc} summarizes our findings.

\section{Radial-velocity analysis and orbital solution}\label{RVs}

As of September 2001, we had in hand a total of 51 radial-velocity
measurements for analysis: 7 from \HIRES\ and 44 from \ELODIE. 
The velocities are all listed in Table~\ref{veltable}.
The
mean uncertainty on the velocities is of the order of 10\,m\,s$^{\rm
-1}$ (systematic error + photon noise) for both instruments. The
\HIRES\ velocities have an arbitrary zero point, and therefore the
orbital solution presented in Table~\ref{orbelem} includes the
velocity offset $\Delta {\rm RV}_{\rm H-E}$ between \HIRES\ and
\ELODIE\ as an additional free parameter. 

The mass given in the table --- $6.29 \pm 0.06$ Jupiter masses (=${\rm
M_{Jup}}$), was derived by assuming a primary mass of $m_1=0.87 \
\mathmsun$ for \hdb\ (T00). The uncertainty of the planet mass is much
larger than the value quoted above, and it depends on the uncertainty of the
primary mass, which was not given in T00. Assuming, for
example, an uncertainty for $m_1$ of 5 per cent, the error on $m_{\rm 2,min}$
would be $0.21 \ {\rm M_{Jup}}$.
Adopting these values for the total mass of
\hdb\ and its planet we get a semimajor axis of about $0.32$
AU. The Hipparcos measurements of \hdb\ are of a very poor quality,
and therefore we have not attempted to use Hipparcos to constrain
the companion mass, in the way done by Zucker \& Mazeh (2001).

\begin{figure}
    \plotone{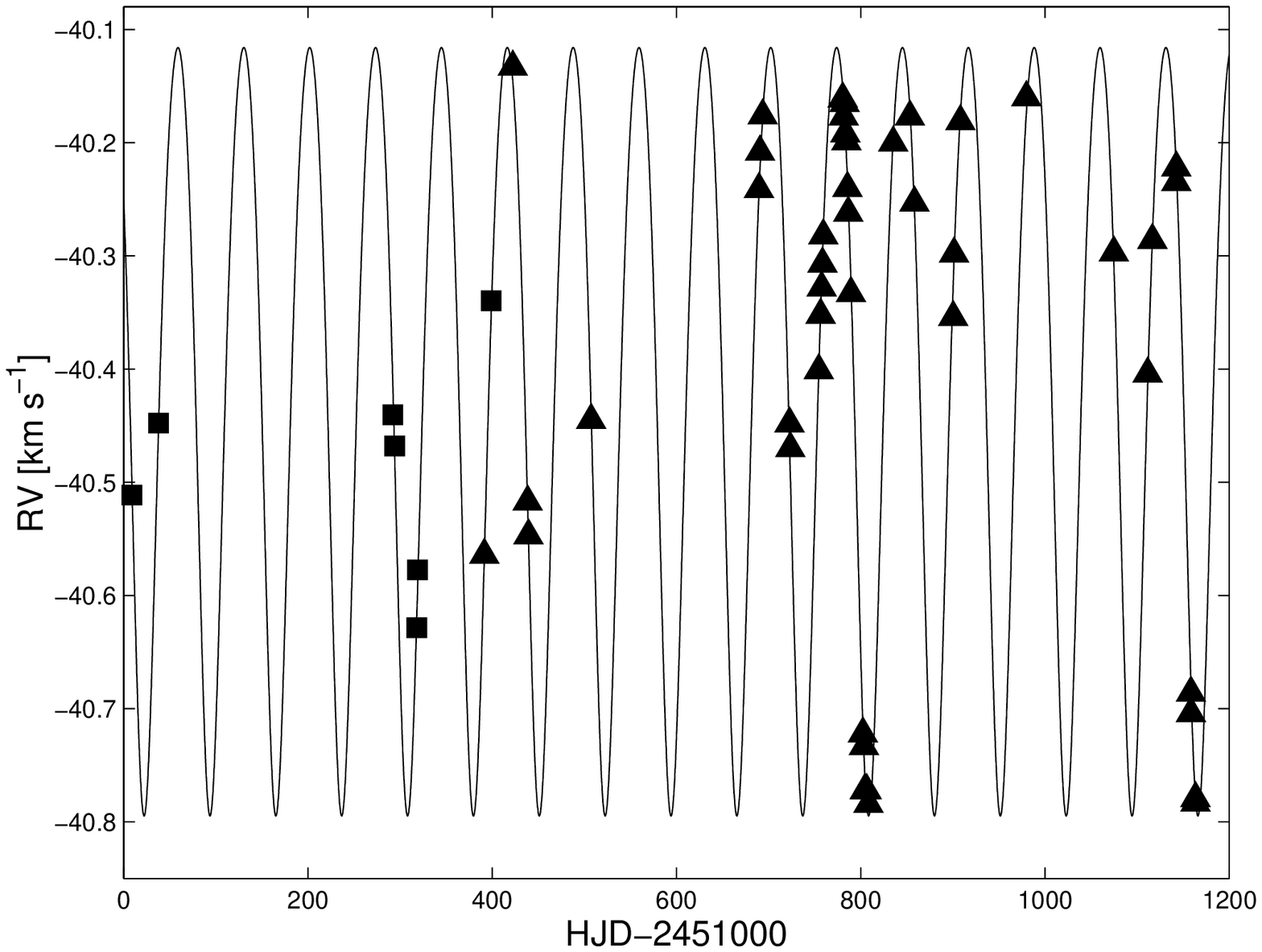}
    \caption{\label{time} \hdb\ radial-velocity data. 
    Triangles: \ELODIE-\OHP\ measurements. 
    Squares: \HIRES-Keck measurements.}
\end{figure}

\begin{figure}
    \plotone{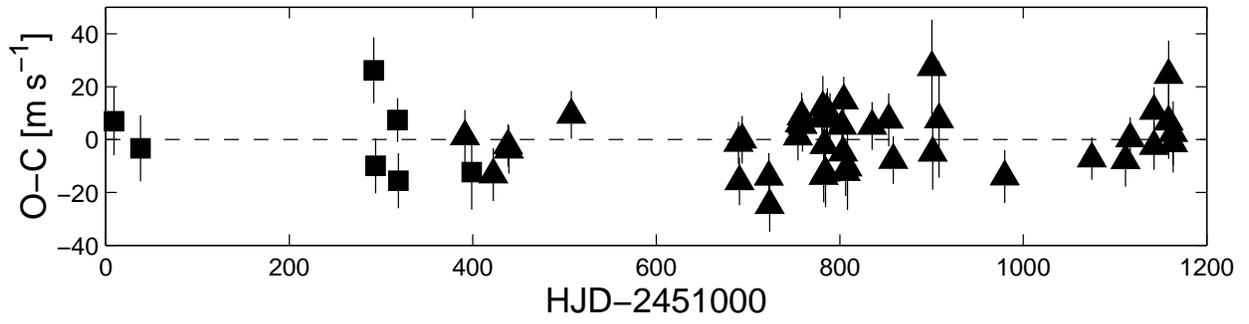}
    \caption{\label{res} The residual radial velocities after subtracting 
      the orbital solution. The error-bars represents the errors of the
      original velocities. Triangles and squares: see Figure~\ref{time}.}
\end{figure}

\begin{figure}
    \plotone{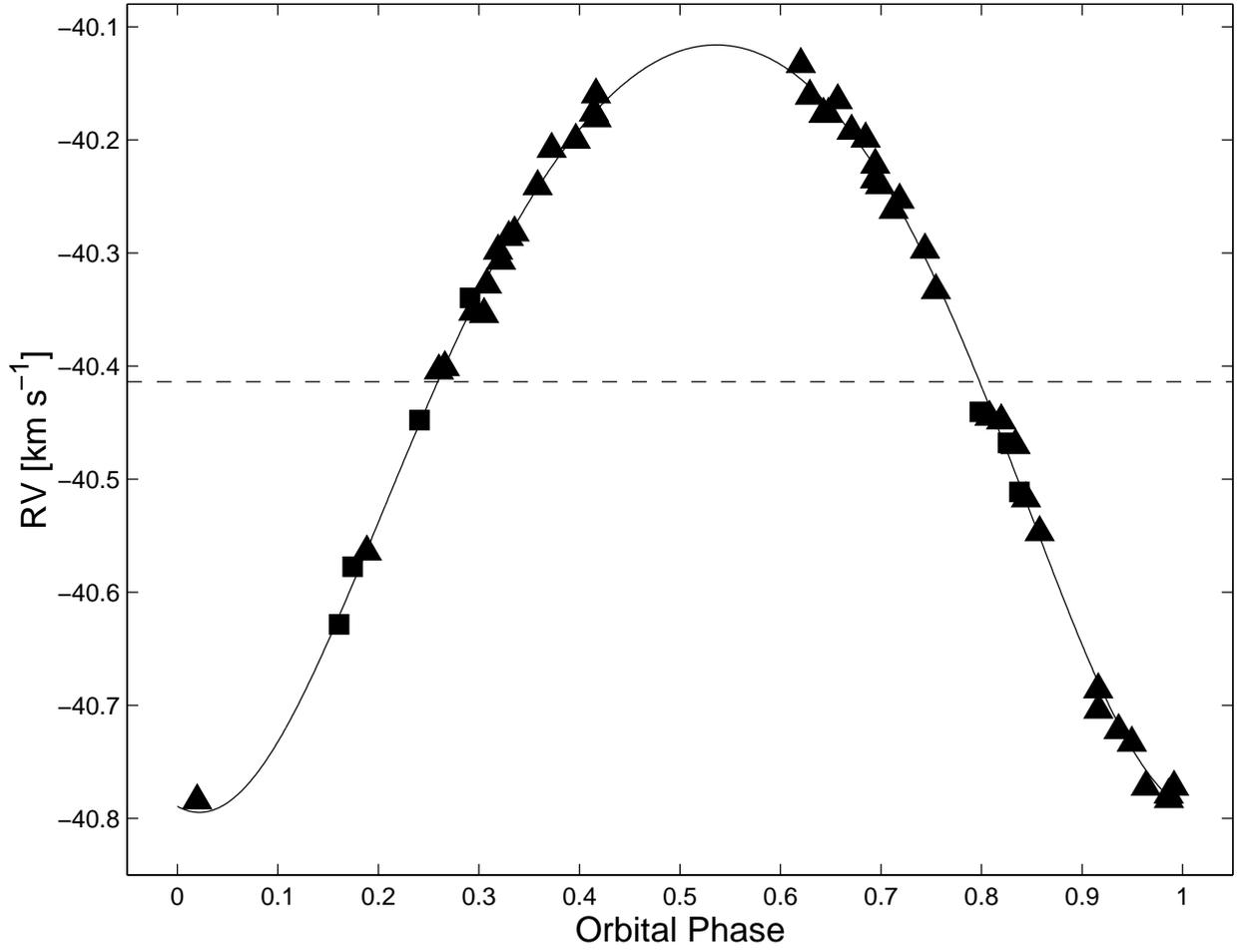}
    \caption{\label{phased} The radial velocities and the orbital fit
      as a function of phase. Triangles and squares: see Figure~\ref{time}.}
\end{figure}

Figure \ref{time} shows the velocities as a function of time after
shifting the \HIRES\ data, and Figure \ref{res} shows the residuals
obtained after subtracting the fitted orbit. The phase-folded
velocities are displayed in Figure \ref{phased}.

\section{Stellar properties of the wide binary} \label{stars}

In this section we first consider previous work on the wide binary
\hd, whose present separation is $16\farcs10 \pm 0\farcs01$ (ESA 1997).  
McAlister et al.\ (1987) resolved the brighter component A by
speckle interferometry into a 0\farcs1 binary, which was subsequently
followed with radial-velocity measurements by T00. The combined
spectroscopic-interferometric orbit of the close pair (T00) yielded a
period of 3.55 years, an eccentricity of 0.59 and a dynamical parallax
of $25\pm 8$ milli-arc-seconds (mas).  Hipparcos (ESA 1997) reported a
somewhat different parallax for \hda\ ---\ $20.4 \pm 1.6$ mas, which
implies a projected separation of $790 \pm 60$ AU. Note, however, that T00
cautioned against the use of the Hipparcos parallax which was computed
without considering the binary nature of A. 

The radial velocity of \hdb\ was measured by T00 to be $-40.65 \pm
0.16$ \kms, while their orbit yielded a center-of-mass velocity for the
close pair A of $-41.01\pm 0.03$ \kms. This similarity, together with
the high common proper motion of A and B, established the physical
association of A and B. Adopting the dynamical parallax we derived a
projected separation of the wide pair of $640 \pm 210$ AU.

Using the photometry by Schoeller et al.\ (1998), T00 concluded that A
comprises a G1-K1 pair, with masses of 1.1 and 0.79 \msun.  Their
model suggested that the B component, around which the planet was
discovered, is of spectral type G8V and a mass of 0.87 \msun.  T00
extracted from spectra of \hd\ some equivalent width measurements,
out of which they derived a solar metallicity for the system. The
metallicity, together with the Galactic velocity of \hd\ led them
to suggest that the system belongs to a disk population of intermediate
age.

We repeated the procedure used for HD~80606 (Naef et
al.\ 2001) and used a high signal--to--noise \HIRES\ spectrum of
\hdb\ to derive its atmospheric parameters (\LTE\ analysis) in the
same way as in Santos et al.\ (2000).  Our final iron line list 
consisted of 16 \ion{Fe}{1} lines and 2 \ion{Fe}{2} lines.  We
estimated the uncertainties on the derived atmospheric parameters in
the same way as in Gonzalez \& Vanture (1998).  The resulting
atmospheric parameters are listed in Table~\ref{atmos}.

Table \ref{atmos} also includes an upper limit for the lithium abundance of \hdb.  To
derive this upper limit we summed all our 44 \ELODIE\ spectra of \hdb\
in the \mbox{$\lambda$ 6707.8 \AA\ \ion{Li}{1}} line region. No trace
of lithium was detected, and therefore a 3$\sigma$ confidence level
upper limit on the equivalent width was derived. The abundance upper
limit was then calculated with the curves of growth by Soderblom et
al. (1993). The lithium abundance was scaled with $\log$\,$n({\rm
H})$\,=\,12. 

\begin{figure}
    \plotone{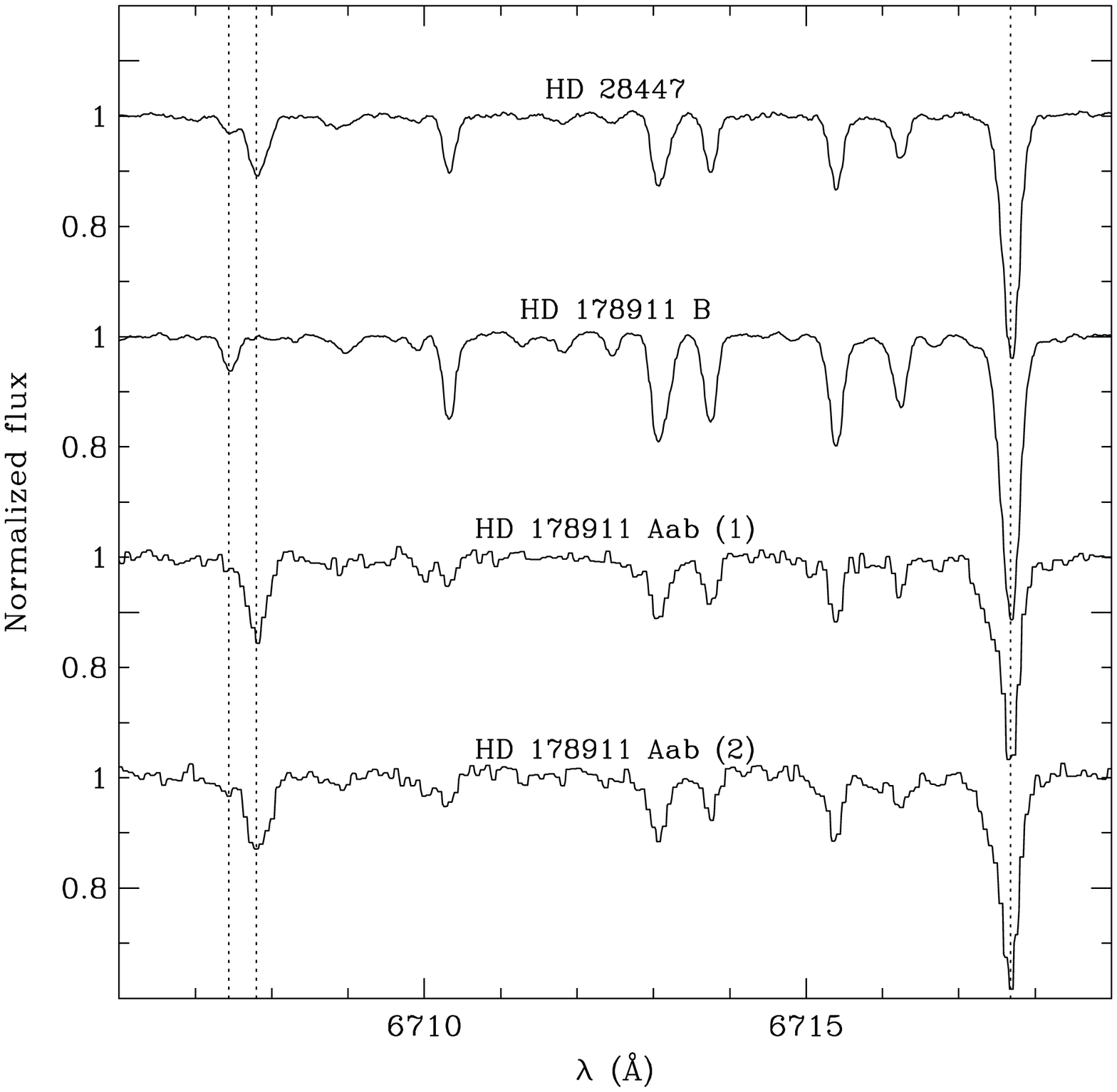}
    \caption{\label{lithiumfig} The \mbox{$\lambda$ 6707.8 \AA\
        \ion{Li}{1}} line region in \ELODIE\ spectra. Three spectral lines
are marked by dotted lines in the Figure: \mbox{$\lambda$ 6707.44 \AA\
\ion{Fe}{1}}, \mbox{$\lambda$ 6707.80 \AA\ \ion{Li}{1}} and
\mbox{$\lambda$ 6717.68 \AA\ \ion{Ca}{1}}.}  
\end{figure}

Interestingly, lithium was definitely detected in \ELODIE\ spectra of
\hda ab.  To show that feature, we plot in Figure \ref{lithiumfig} the
relevant region of two spectra of \hda ab, the co-added spectra of
\hdb, and a spectrum of a G5 comparison star with lithium.  The two
spectra of \hda ab were shifted according to the calculated velocity of the
massive component, to align the relevant lines with those in the other
spectra.  One can note the absence of the \ion{Li}{1} line in \hdb,
and the clear presence of this line in \hda. The asymmetry of the
\ion{Ca}{1} line demonstrates the binary nature of \hda, as the
fainter component is shifted relative to the brighter one by 0.29 \AA\
and 0.2 \AA\ in the two spectra.  This blending of the lines of the
two components of the close binary A did not allow us to perform a
quantitative analysis of the lithium abundance of A. However, we
adopted the value 50\,m\AA\ as a very rough estimate for the
equivalent width of the lithium line, 
with a probable uncertainty of 10\,m\AA. 
This value, together with a
\Teff\ estimate of 5910~K (Feltzing \& Gustafsson 1998), implies
$\log$\,$n({\rm Li})\,=\,2.4$ for the Aa component, with a probable
error
of 0.1.

\section{The Lithium Abundance Difference of \hd --- a Comparison with
16~Cyg} \label{lithium}

In this section we discuss the difference between the lithium
abundance of \hda\ and B. This difference is intriguing, as it reminds
us of a similar difference between 16~Cyg~A and B --- a wide binary in
which a planet was found to orbit around the B component (Cochran et
al.\ 1997).  Another intriguing similarity between the case of 16~Cyg
and \hd\ is hinted by the detection of a faint M-star companion to
16~Cyg~A by Hauser \& Marcy (1999), at a separation of
3\farcs2. Turner et al. (2001) suggested the companion --- 16~Cyg~Ab
--- is not physically associated with 16~Cyg~Aa, but Lloyd et
al. (2001) recently claimed that its measured proper motion indicates
that the Aa and Ab components are bound. Assuming this is the case,
the A component of 16~Cyg, like the A component of \hd, is a
binary, with a projected separation of about 80 AU.  We
summarize below the explanations suggested for the lithium abundance
pattern in the case of 16~Cyg and then apply them to \hd.

The two components of the 16~Cyg wide binary are known to be almost
identical stars, with a temperature difference smaller than 50~K
(Friel et al. 1993). 
Naively, one would expect them to have similar
lithium abundances. However, it was found (Friel et al.\ 1993; King et
al.\ 1997) that the lithium abundance of 16~Cyg~A is larger by at
least a factor of 5 than that of B.  

In general, photospheric lithium
abundance in cool stars is a useful stellar evolution model
diagnostic because this element is destroyed at temperatures of a few
million degrees at the base of the convective zone. Although our
understanding of lithium abundance evolution is far from complete,
present models include burning of lithium during the early
pre-main-sequence (PMS) phase of evolution, probably together with
mixing material slowly over the lifetime of the star in the radiative
layers below the convection zone (e.g., Ryan \& Deliyannis 1995;
Deliyannis et al.\ 2000).  Therefore, the difference in the lithium
abundance between the A and B components of the 16~Cyg wide binary
attracted attention even before the discovery of the planet (Friel et
al. 1993). Since the discovery of the planet around 16~Cyg~B in 1997
quite a few studies suggested posssible explanations for this
difference.

King et al.\ (1997) commented that ``it may in principle be possible
for planets or associated circumstellar disks to affect a parent
star's initial angular momentum and/or its subsequent evolution and
thus induce or inhibit internal mixing resulting in lithium depletion.''
Cochran et al. (1997), in their planet discovery paper, were more
definite in their effort to explain the low lithium abundance of 16~Cyg~B. 
They pointed out that the angular momentum history of young
solar-type stars is governed strongly by torques exerted on the star
by the inner accretion disk. The general wisdom is that PMS stars with
massive disks exhibit slow rotation rates resulting from the magnetic coupling 
to the inner disk (e.g. Choi \& Herbst 1996; but see
Stassun et al.\ 1999 for a dissenting view). Slow rotators are known to
have low lithium abundance (e.g., Soderblom et al. 1993).  The fact that
only 16~Cyg~B was found to have a planet and not the A component
suggested, according to Cochran et al., that B had a more massive
disk, out of which the planet was formed. The same massive disk
induced slowing down of the stellar rotation at the PMS stage and
therefore led to lithium depletion in its atmosphere.

Some support for this scenario has been presented by Gonzalez \&
Laws (2000). They pointed out that when corrected for differences
in temperature, metallicity and chromospheric activity, the lithium
abundances of planet-hosting stars are in general lower than normal
stars. The evidence for this general trend was disputed by Ryan
(2000), who claimed that the lithium abundances of planet-harboring
stars are indistinguishable from those of non-planet-harboring stars
of the same age, temperature and composition. He further argued that
the small difference in temperature between 16~Cyg~A and B (Deliyannis
et al. 2000; Laws \& Gonzalez 2001) could explain the lithium
abundance difference, although the Li-\Teff\ slope might be
somewhat steeper in the 16~Cyg system than observed elsewhere. From
this point of view the planet around B has nothing to do with the
lithium abundance difference.

A completely different explanation for the lithium abundance
difference between 16~Cyg~A and B was suggested by Gonzalez (1998). In
a paper where he invoked a pollution scenario for the {\it high}
metallicity of the stars hosting planets, he suggested that 16~Cyg~A
ingested a planet that was pushed into its atmosphere by tidal
interaction with the existing planet of the B component. The swallowed
planet raised the lithium abundance of 16~Cyg~A dramatically. Laws \&
Gonzalez (2001) found some support for this scenario by their finding
that the iron abundance of 16~Cyg~A is slightly higher than B. They
argued that both the slightly higher metallicity and the large
difference in the lithium abundance could be attributed to a scenario in
which A swallowed a giant planet into its atmosphere. 
Evidence for
such a process where 
a planet was swallowed by its host star was also found in the case
of HD~82943 (Israelian et al. 2001).

To summarize, three explanations have been put forward to explain the
lithium abundance difference between 16~Cyg~A and B, two of which had
to do with the fact that B has a planet around it, while A does
not. One of the two attributed the depletion of lithium in B to the
massive disk that eventually formed the planet, while the other
attributed the relatively high lithium abundance of A to the ingestion
of a planet by A. The third explanation was based on the temperature
difference between A and B.

We can now examine the relevance of the three explanations to the \hd\
system. Here also the lithium depletion was found in the
planet-hosting star which is also the cooler star. The low lithium
abundance of B could be the result of its massive disk, as the first
explanation suggests.  The lithium abundance difference
might also be
the result of the different temperatures of A and B.  The value we
got for the lithium abundance of B, together with the rough estimate
for A, suggest a difference of at least 1.6 in
$\log$\,$n({\rm Li})$.  The calibration of Gonzalez \& Laws (2000)
predicts a much smaller difference of 0.44 for the same \Teff\
difference.  However, the explanation suggested by Ryan (2000) for
16~Cyg might apply here also.  
If we extrapolate his suggestion of a 0.33 dex decline in
$\log$\,$n({\rm Li})$ per 50~K, we get a difference of 1.72 dex for the
HD~178911 system, consistent with our estimate. Then again, it
is not clear whether the slope suggested by Ryan is applicable 
to the whole range of temperatures covered by the two components of 
\hd.

When we try to invoke the planet-ingestion scenario to produce
a lithium enhancement in \hda, we face some difficulties, because of the 
relative proximity, 3 AU,  of the Aa and Ab components. 
The tidal forces of Ab would have disrupted
any planetary orbit
around Aa in a very early stage, especially under the migration
hypothesis which requires the planet to form in a distance of a few AU
from its parent star.
However, to still hold to this scenario for \hd\, we might try to invoke the
formation of a circumbinary planet. Such a scenario is possible,
although no circumbinary planet has been discovered yet (e.g. Holman
\& Wiegert 1999).
This hypothetical planet could have migrated inwards through the standard
migration mechanism, or interact with the planet of
\hdb\ in the way suggested for 16~Cyg by Gonzalez (1998).
At close enough distance to the binary, the resulting instability
might lead to ingestion of the planet by one of the binary components.
We will not face the same difficulties in 16~Cyg~A,
even if we accept the triple nature of 16~Cyg, 
since the 16~Cyg~A pair is wide enough (80~AU) to allow
the formation and evolution of a planet around Aa.
Thus, in principle, we cannot completely rule out
any of the three explanations, although the
planet-ingestion seems less likely for \hd\ as it requires some non-standard
formation processes.

Interestingly, Ryan \& Deliyannis (1995) suggested that binaries with
sufficiently short periods might have preserved more lithium than
normal, single stars.  They relied on the theory of Zahn and Bouchet
(1989; but see the different approach by Mathieu and Mazeh 1988 and
Goldman and Mazeh 1991) that short-period binaries, with periods
shorter than 8--9 days, reached synchronization during the {\it early}
stages of the PMS phase.  At this early fully convective stage, the
stellar interiors were not hot enough to destroy lithium. The depletion
of lithium occured in a later stage at the stellar interior by
``angular momentum transport mixing''. Short-period tidally-locked
binaries did not go through this mixing stage because of their
different angular momentum evolution and therefore display high
lithium abundance. Whatever the mechanism would be, it seems that the
long period of the \hda\ binary renders this argument inapplicable to
our system.

\section{Conclusion} \label{conc}

We have shown evidence for a planet orbiting \hdb, at a separation of
about 0.3 AU. 
\hdb\ and 16~Cyg~B are currently the only triple systems known to
harbor planets.
In both cases the planet was discovered around the distant component,
while the other two stars comprise a close binary.

Can there be another planet around one of the components of \hda?
According to the presently accepted paradigm, the
short period planets have migrated from where they had formed, at a
few AU (but see, e.g., Boss 1998 for an alternative). 
At such large distances, no planet could have survived the
tidal interaction of the other star in the close pair \hda. 
It seems that under the present paradigm the only possible planets 
in the close pair are circumbinary planets. 

Assuming that at present there is no planet around \hda, it is interesting to consider
the formation of the whole \hd\ system. Suppose that the visual binary
was formed by a fragmentation event during the collapse of a molecular
cloud core (e.g., Burkert \& Bodenheimer 1993).  The cloud core
collapsed until it was of the order of a thousand AU in diameter, at
which time it fragmented into two parts. One part ended up as the
close stellar pair A, while the other one formed (at least) one planet
which we have discovered. It would be interesting to figure out what
were the differences in the two cores that caused them to evolve on
such different formation tracks, and whether the lithium abundance
of A and B can shed some light on our understanding of their
evolution.

\begin{acknowledgements}
We acknowledge support from the Swiss National Research Found 
({\footnotesize FNRS}), the Geneva University and the French 
{\footnotesize CNRS}, the US-Israel Binational Science Foundation
through grant 97-00460 and the Israeli Science Foundation (grant
no.\ 40/00).
 We are grateful to the Observatoire de 
Haute-Provence for the generous time allocation. 
This research has made use 
of the {\footnotesize SIMBAD} database, operated at 
{\footnotesize CDS}, Strasbourg, France.
\end{acknowledgements}

\clearpage
\begin{deluxetable}{cccc}
\tablecaption{\label{veltable}}
\tablewidth{0pt}
\tablecolumns{4}
\tablehead{
\colhead{JD-2400000} & 
\colhead{RV} & 
\colhead{Error} &
\colhead{Instrument} \\
\colhead{} &
\colhead{\kms} &
\colhead{\kms} &
\colhead{}}
\startdata
51008.976 & $-$40.511 & 0.013 & H \\
51037.797 & $-$40.448 & 0.012 & H \\
51292.098 & $-$40.440 & 0.012 & H \\ 
51294.111 & $-$40.468 & 0.010 & H \\ 
51318.033 & $-$40.628 & 0.008 & H \\
51318.992 & $-$40.577 & 0.010 & H \\
51391.487 & $-$40.564 & 0.010 & E \\
51398.821 & $-$40.340 & 0.014 & H \\
51422.356 & $-$40.133 & 0.010 & E \\
51438.359 & $-$40.517 & 0.008 & E \\
51439.325 & $-$40.547 & 0.009 & E \\
51507.253 & $-$40.445 & 0.009 & E \\
51689.592 & $-$40.241 & 0.008 & E \\
51690.580 & $-$40.208 & 0.009 & E \\
51693.587 & $-$40.176 & 0.009 & E \\
51722.547 & $-$40.448 & 0.009 & E \\
51723.560 & $-$40.470 & 0.010 & E \\
51754.470 & $-$40.401 & 0.009 & E \\
51756.498 & $-$40.352 & 0.008 & E \\
51757.458 & $-$40.328 & 0.008 & E \\
51758.456 & $-$40.307 & 0.009 & E \\
51759.424 & $-$40.282 & 0.010 & E \\
51780.438 & $-$40.161 & 0.009 & E \\
51781.406 & $-$40.177 & 0.012 & E \\
51782.405 & $-$40.165 & 0.010 & E \\
51783.394 & $-$40.192 & 0.009 & E \\
51784.393 & $-$40.199 & 0.013 & E \\
51785.421 & $-$40.240 & 0.009 & E \\
51786.393 & $-$40.262 & 0.009 & E \\
51789.396 & $-$40.333 & 0.009 & E \\
51802.384 & $-$40.722 & 0.009 & E \\
51803.323 & $-$40.733 & 0.009 & E \\
51804.329 & $-$40.772 & 0.009 & E \\
51806.327 & $-$40.772 & 0.009 & E \\
51808.347 & $-$40.784 & 0.016 & E \\
51835.272 & $-$40.200 & 0.009 & E \\
51853.240 & $-$40.177 & 0.010 & E \\
51858.294 & $-$40.253 & 0.009 & E \\
51900.241 & $-$40.354 & 0.018 & E \\
51901.228 & $-$40.298 & 0.014 & E \\
51908.244 & $-$40.181 & 0.022 & E \\
51979.684 & $-$40.160 & 0.010 & E \\
52074.570 & $-$40.297 & 0.008 & E \\
52111.478 & $-$40.404 & 0.010 & E \\
52116.452 & $-$40.286 & 0.008 & E \\
52142.501 & $-$40.235 & 0.009 & E \\
52142.515 & $-$40.222 & 0.009 & E \\
52158.367 & $-$40.686 & 0.014 & E \\
52158.380 & $-$40.704 & 0.013 & E \\
52163.387 & $-$40.779 & 0.011 & E \\
52163.400 & $-$40.783 & 0.012 & E \\
\enddata
\tablecomments{H: \HIRES, E: \ELODIE. The \HIRES\ velocities were
  shifted to the \ELODIE\ zero-point.}
\end{deluxetable}

\clearpage
\begin{deluxetable}{llr@{  \,$\pm$\,  }l}
\tablecaption{\label{orbelem} Best-fit orbital solution}
\tablewidth{0pt}
\tablehead{}
\startdata
$P$                         & (days)                           & 71.487     &
0.018\\
$T$                         & (JD)\tablenotemark{\dagger}      & 50\,305.70 &
0.62\\
$e$                         &                                  & 0.1243     &
0.0075\\
$\gamma$                    & (km\,s$^{\rm -1}$)               & $-$40.4138   &
0.0018\\
$w$                         & ($\degr$)                        & 169.8      &
3.6\\ 
$K_{\rm 1}$                 & (m\,s$^{\rm -1}$)                & 339.3      &
3.1\\
$\Delta {\rm RV}_{\rm H-E}$ & (km\,s$^{\rm -1}$)               & $-$40.4886       
& 0.0053\\
$a_{\rm 1} \sin i$          & ($10^{\rm -3}$AU)                & 2.212      &
0.021\\
$f_{\rm 1}(m)$              & ($10^{-9}$${\mathrm M_{\odot}}$) & 2.826      &
0.079\\
$m_{\rm 2,min}$             & (${\mathrm M_{\rm Jup}}$)        & 6.292      &
0.059\\
\noalign{\vspace{0.025cm}}
\hline
\noalign{\vspace{0.025cm}}
$N$                         &                                  &
\multicolumn{2}{c}{44(E) + 7(H)}\\ 
$\sigma_{\rm O-C}$          & (m\,s$^{\rm -1}$)                &
\multicolumn{2}{c}{11.0 (E:10.6, H:13.6)}\\
\enddata
\\
\tablenotetext{\dagger}{JD\,=\,HJD\,$-$\,2\,400\,000}
\end{deluxetable}

\clearpage
\begin{deluxetable}{lcr@{$\,\pm\,$}l}
\tablecaption{\label{atmos} The measured atmospheric parameters of
  \hdb}
\tablewidth{0pt}
\tablecolumns{4}
\tablehead{}
\startdata
\Teff                 & (K)     & 5650   & 80 \\
$\log g$              & (cgs)   & 4.65   & 0.15 \\
$\xi_{\rm t}$         & (\kms)  & 0.85   & 0.19 \\
$[$Fe/H$]$            &         & 0.28   & 0.08 \\
$W_\lambda({\rm Li})$ & (m\AA)  &\multicolumn{2}{c}{$<2.3$} \\
$\log n({\rm Li})$    &         & \multicolumn{2}{c}{$<0.75$} \\
\enddata
\end{deluxetable}

\end{document}